\documentclass[twocolumn,prl,superscriptaddress,aps,amsmath,floatfix,longbibliography]{revtex4-2}

\usepackage{graphics}
\usepackage{graphicx}
\usepackage{epstopdf}
\usepackage{dcolumn}
\usepackage{bm}
\usepackage{longtable}
\usepackage{epsfig}
\usepackage{times}
\usepackage{url}

\usepackage{color}

\begin{document}
\title{Gravitationally sensitive structured x-ray optics using nuclear resonances}
\author{Shin-Yu   \surname{Lee}}
\affiliation{Department of Physics, National Central University, Taoyuan City 32001, Taiwan}

\author{Sven \surname{Ahrens}}
\affiliation{Shanghai Normal University, Shanghai 200234, China} 

\author{Wen-Te \surname{Liao}}
\email{wente.liao@g.ncu.edu.tw}
\affiliation{Department of Physics, National Central University, Taoyuan City 32001, Taiwan}
\affiliation{Physics Division, National Center for Theoretical Sciences, Taipei 10617, Taiwan}
\affiliation{Center for Quantum Technology,  Hsinchu 30013, Taiwan}
\date{\today}
\begin{abstract}
Einstein's general theory of relativity not only revolutionized human understanding of the universe, but also brought many  gravitational applications in large scale, such as gravitational-wave astronomy \cite{GW150914}, gravitational lensing \cite{Bartelmann2001}, and the operation of the global positioning system \cite{Ashby2002}.
However, it still remains a challenge to implement applications for gravitational effects at  small spacial extensions on Earth.
Here, we investigate a  structured waveguide system that  allows for the control of an x-ray profile at altitude separations of millimeters and even shorter using the nuclear resonant scattering of x rays \cite{Hastings1991,Roehlsberger2004}.
Our present results suggest a potential compact scheme for turning the Earth's gravity  into a practical application of x-ray optics.
\end{abstract}
\maketitle
The Pound–Rebka experiment \cite{Pound1960} has demonstrated a unique system for probing the gravitational red-shift effect by exploiting an extremely narrow nuclear linewidth in combination of a high x-ray energy in the M\"ossbauer effect \cite{Kalvius2012}.
Moreover, advances in modern x-ray light sources and optics have raised the field of x-ray-nuclei interactions to a new level of accuracy where coherent quantum control comes into play \cite{Shvydko1996, Shvydko1999N, Palffy2009, Roehlsberger2012, Liao2012a, Adams2013, Vagizov2014, Liao2015, Heeg2015a, Heeg2017, Zhang2019, Radeonychev2020,Heeg2021,Chen2022}.
A combination of  nuclear quantum coherence and its sensitivity to gravity will potentially lead to a new type of x-ray optics whose performance depends on the gravitational red-shift in addition to the typical Zeeman shift \cite{Shvydko1996,Liao2012a} and Doppler shift \cite{Vagizov2014,Heeg2017, Zhang2019,Radeonychev2020}.
In this context, we investigate a system that emulates the Schr\"odinger equation \cite{Marte1997} and is sensitive to gravity.
The present scheme allows for a systematic generation of structured x rays \cite{Seipt2014, Rubinsztein2016,  Macdonald2017, Forbes2021} via changing the altitude, the x-ray photon energy, or the external magnetic field of a our system for a given waveguide structure.
%
%
%
\begin{center}
\begin{table*}[b]
\setlength{\tabcolsep}{6pt}
\vspace{-0.4cm}
\caption{\label{table1}
{\bf Nuclear and waveguide material parameters}. 
For each isotope $X$ we present 
the nuclear transition energy $E_t$ and
the radiative decay rate $\Gamma$.   
The last six columns list the  x-ray index of refraction $n_e = 1-\delta+i \beta$ for  SXWG  materials  \cite{Roehlsberger2004, nndc, lbl}.
}
\begin{tabular}{rrclcrrrrrrr}
\hline
\hline
$X$ &   $E_{t}$ (keV) & $\Gamma$ (MHz) &   $\delta_X  (10^{-6})$ & $\delta_\mathrm{C}  (10^{-6})$ &  $\delta_\mathrm{Pt}  (10^{-5})$ &  $\beta_X (10^{-9})$ & $\beta_\mathrm{C} (10^{-9})$  & $\beta_\mathrm{Pt} (10^{-6})$
\\

\hline
$^{45}$Sc   & 12.4   &  2.18$\times 10^{-6}$  &     3.84 &  2.97  & 2.091 &  131.9  &  1.78 &  2.737\\

$^{57}$Fe   & 14.413 &  7.05                  &     7.43 &  2.20  & 1.607 &  338.9  &  0.93 &  2.49 \\

$^{73}$Ge   & 13.275 &  0.24                  &     5.41 &  2.59  & 1.622 &  508.9  &  1.32 &  2.947\\

$^{181}$Ta  &  6.238 &  0.11                  &    67.74 & 11.77  & 8.731 & 7987.3  & 32.76 & 12.616\\

$^{182}$Ta  & 16.273 &  2.45$\times 10^{-6}$  &    10.41 &  1.72  & 1.304 & 1062.2  &  0.55 &  1.642\\
\hline
\hline
\end{tabular}
\end{table*}
\end{center}
\begin{figure*}[b]
\includegraphics[width=1\textwidth]{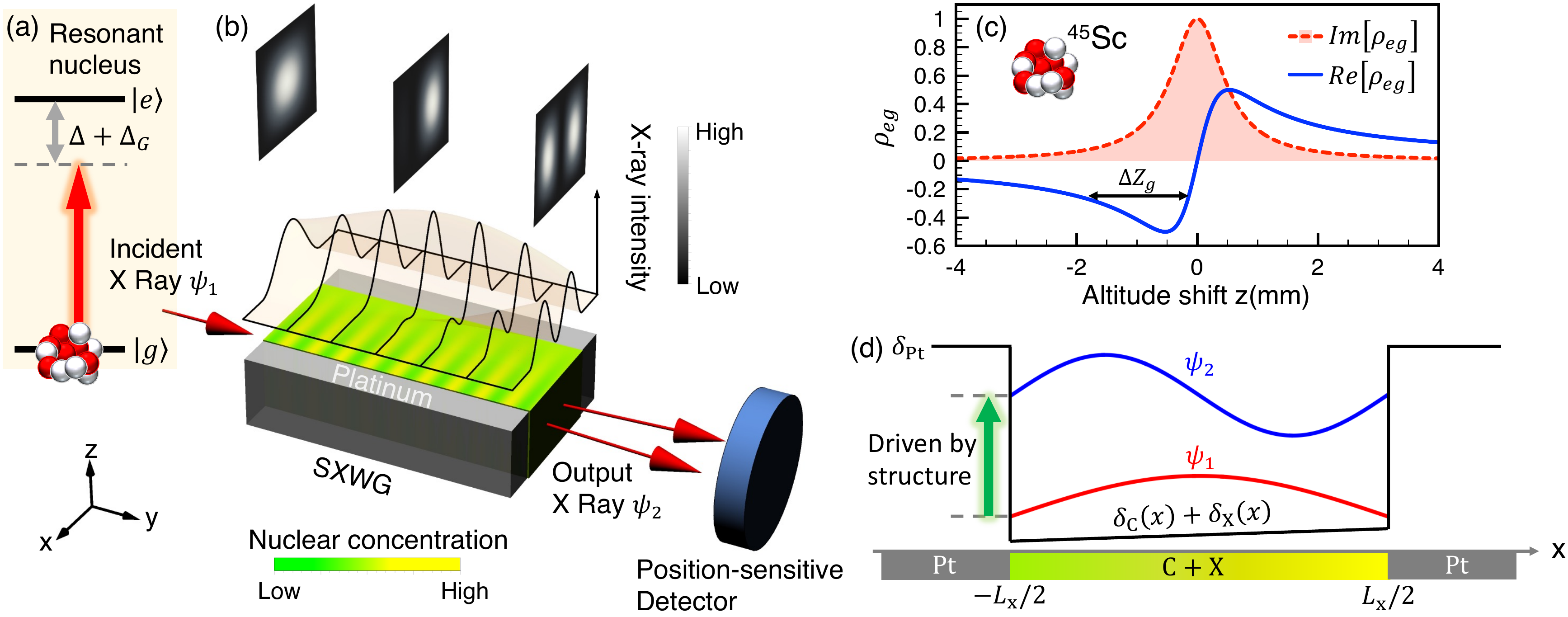}
\caption{\label{fig1} 
(a) the incident x ray (red upward arrow) drives a nuclear transition $\vert g\rangle \rightarrow \vert e\rangle$ with detuning $\Delta+\Delta_G$ (gray vertical double arrows).
$\Delta$ is the x-ray detuning, and $\Delta_G$ is the x-ray gravitational red shift.
(b) a hard x ray of transverse mode $\psi_1$ (red arrow) propagates through a structured platinum cladding waveguide with a periodic nuclear distribution whose  spatial concentration is indicated by the horizontal green-yellow legend. An output x ray of mode $\psi_2$ is measured by a downstream position-sensitive detector (blue thick disc).
The top black curves and pictures in gray level represent the intra-waveguide x-ray intensity.
(c) the SXWG altitude-dependent nuclear coherence $\rho_{eg}$ between the nuclear ground state $\vert g\rangle$ and the excited state $\vert e\rangle$ of the isotope $^{45}$Sc. Red-dashed (blue-solid) line depicts the imaginary (real) part of the nuclear $\rho_{eg}$. 
Black horizontal double arrow indicates  the   full altitude width $\Delta Z_G$  at the half maximum $\vert Re\left[ \rho_{eg}\right] \vert$.
(d) the transversely gradient (along x) and the longitudinally periodic (along y) electronic refractive index of an SXWG. 
Between the platinum claddings the intra-waveguide structure, made of carbon and an isotope $\mathrm{X}$, drives the transition from the x-ray ground state $\psi_1$ (red-solid line) to the first excited state $\psi_2$ (blue-solid line).
}
\end{figure*}
%
\begin{figure}[b]
\includegraphics[width=0.45\textwidth]{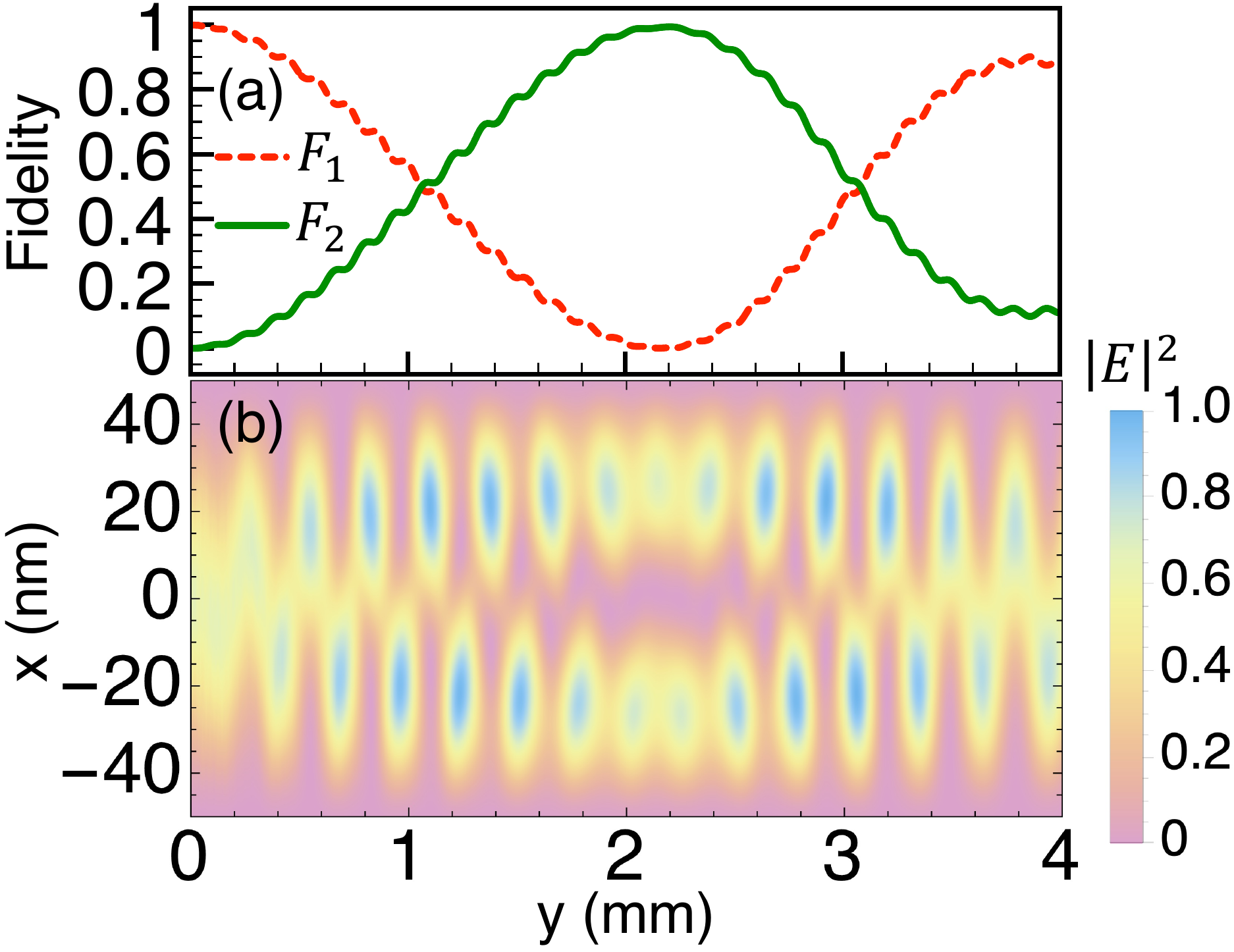}
\caption{\label{fig2}
A full cycle of the x-ray Rabi oscillation between  $\psi_1$ and $\psi_2$  is illustrated by
(a) the fidelity $F_1$ (red-dashed line) and $F_2$ (green-solid line) and 
(b) the normalized intra-waveguide x-ray intensity distribution.
The input x-ray in  the state $\psi_1$  splits and reaches the maximum transverse double-hump separation of the state  $\psi_2$  at $y=2$mm  where the maximum $F_2$ also occurs. 
For $y>2$mm the x-ray confluence reflects  the second half Rabi cycle, and the transverse x-ray pattern returns toward the state $\psi_1$.
}
\end{figure}
%
%
%
%
\begin{figure*}[t]
\includegraphics[width=1\textwidth]{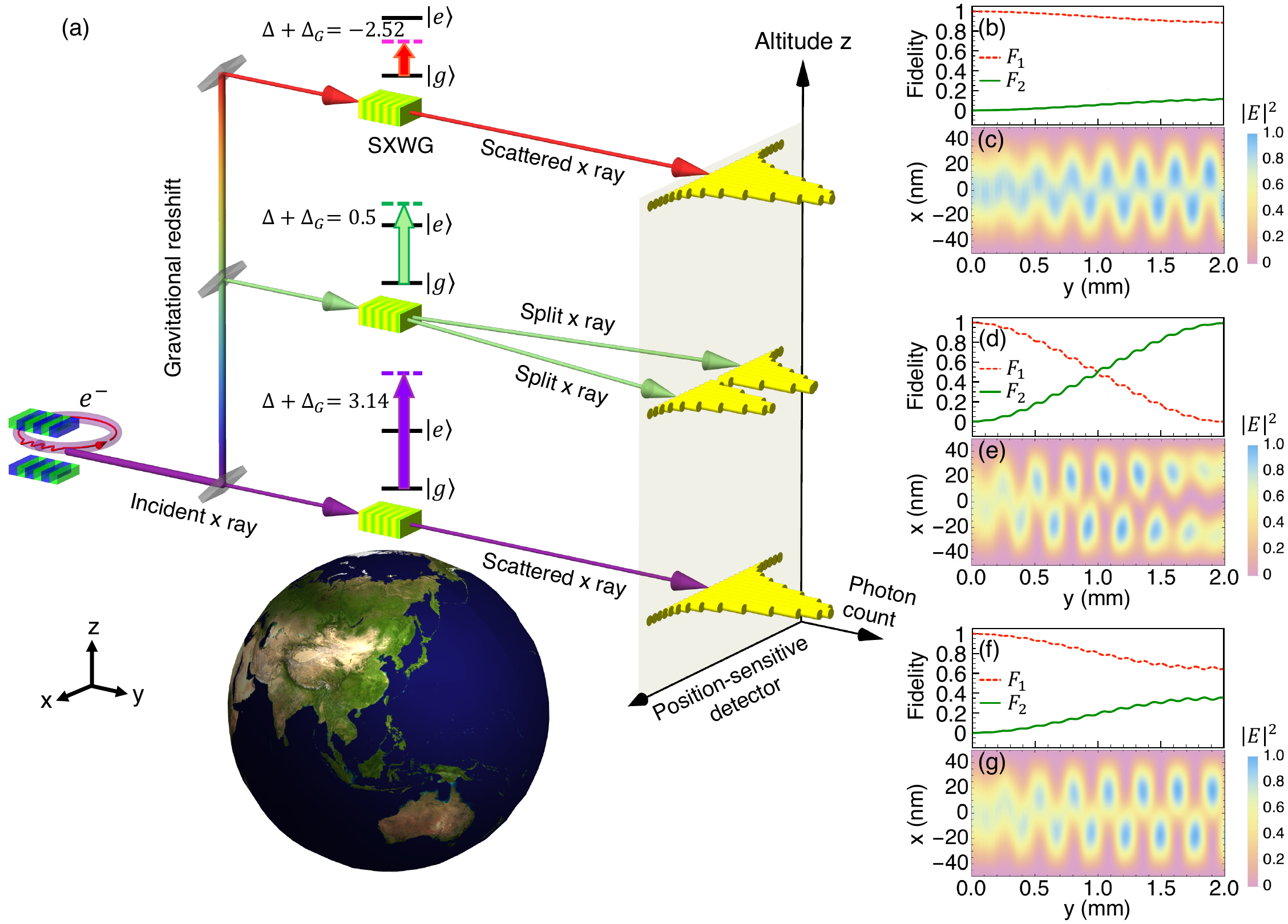}
\caption{\label{fig3} 
(a) Earth's gravity changes the x-ray propagation in the waveguide composed of $^{45}$Sc nuclei.  Three cases at different altitudes $z=2.32$cm,  $z=2$cm, and $z=1.72$cm  where x rays propagate with a  detuning   $\Delta = 19.36$ and gravitational red shifts $\Delta_G=-21.88$, $\Delta_G=-18.86$, and $\Delta_G=-16.22$, respectively.
(b,  d,  and f) the fidelity $F_1$ (red-dashed line) and $F_2$ (green-solid line) for $z=2.32$cm, $z=2$cm, and $z=1.72$cm,  respectively. 
(c,  e,  and g) the normalized intra-waveguide x-ray intensity distribution $\vert E \left( x,y \right) \vert^2$ at altitude $z=2.32$cm, $z=2$cm, and $z=1.72$cm  from the top to the bottom. 
}
%
%
\includegraphics[width=1\textwidth]{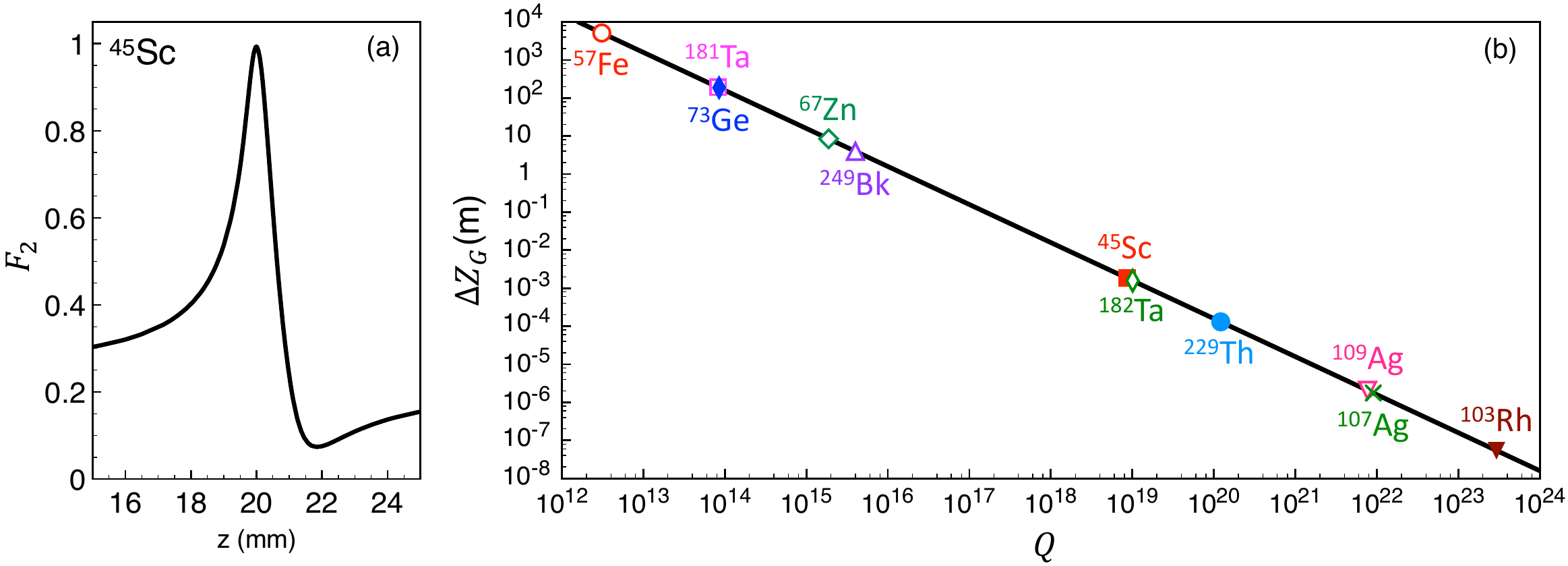}
\caption{\label{fig4}
(a) altitude-dependent x-ray fidelity $F_2$ through an SXWG composed of $^{45}$Sc nuclei.
(b) the FWHM $\Delta Z_G$ on  Earth is dependent on the quality factor $Q$ of nuclear resonances for different nuclear species.
}
\end{figure*}

The system is depicted in Figure~\ref{fig1} with the description as follows. 
An x ray drives a nuclear transition from the ground state $\vert g\rangle$ to the excited state $\vert e\rangle$ with the total detuning $ \Delta_t \Gamma = \left(  \Delta_G+\Delta\right) \Gamma $ in Fig.~\ref{fig1}(a).
We emphasize that the gravitational red shift $\Delta_{G}\simeq  -z E_t G M_E / \left( \hbar\Gamma c^2 R_E^2\right)  $ has to be taken into account when the system is located at different altitude $z$ relative to where x rays are emitted.
Here
$E_t$ is the nuclear transition energy,  
$G$ is the gravitational constant,
$M_E$ is the mass of the Earth, and
$R_E$ is the average radius of the Earth.
$\Delta\Gamma$ is the x-ray detuning, and $\Gamma$ the spontaneous decay rate of the excited state $\vert e\rangle$.
Our system of nuclear resonant scattering of x rays can be described by the optical-Bloch equation \cite{Shvydko1999N,Chen2022}:
\begin{eqnarray}
\partial_{t}{\rho_{eg} } &=& -\Gamma\left[\frac{1}{2}-i\left( \Delta + \Delta_G \right)\right]  \rho_{eg}+\frac{i}{2}\frac{P}{\hbar}E ,\label{eq1} \\
\frac{1}{c}\partial_{t} E + \partial_{y} E &=& \frac{-1}{2ik} \nabla_{\perp}^2 E -\frac{k}{2i}(n_e^2-1)E +i\eta\rho_{eg}  , \label{eq2}
\end{eqnarray}
where $\rho_{eg} $  is the  coherence of a nuclear two-level system, $E$ is the x-ray electric field strength,  $\hbar$ is the reduced Planck constant, $k$ is the x-ray wavenumber, and $n_e$ is the index of refraction from electrons.
Further, we denote the transverse Laplacian $\nabla_{\perp}^2 = \partial^2_x+\partial^2_z$, and the coupling constant $\eta=2\hbar\Gamma\xi/\left( P L\right) $, where $L$ the length of the waveguide,  $P$ is the nuclear transition dipole moment,  and $\xi$ is the nuclear resonant thickness. 
The steady state of Eqs.~(\ref{eq1}-\ref{eq2}) leads to the analytic solution  
$\rho_{eg} = P E \left( i - 2\Delta_t   \right) / \left[\hbar\Gamma\left( 1+4\Delta_t^2 \right) \right] $ and the optical Schr\"odinger equation  \cite{Marte1997} (see supplemental information)
\begin{equation}\label{eq4}
i\hbar c\partial_{y} E =  -\frac{\hbar^2}{2m_e}  \nabla_{\perp}^2 E  + \hbar c \left[ \frac{ k (1-n_e^2)}{2}
+\frac{4\xi \Delta_t }{L \left( 1+4\Delta_t^2 \right)} \right]  E,
\end{equation}
with the effective mass $ m_e = \hbar k/c$. 
Equation \eqref{eq4} suggests  a structured x-ray waveguide (SXWG) system composed of resonant nuclei, as depicted in Fig.~\ref{fig1}(b), with a high degree  of freedom for simulating different quantum systems via spatial  engineering  of $n_e$ and $\xi$.
Fig.~\ref{fig1}(c) displays  the altitude dependent real part $Re\left[ \rho_{eg}\right] $ (blue-solid) and  imaginary part $Im\left[ \rho_{eg}\right] $ (red-dashed line)  of  $\rho_{eg}$ with $\Delta = 0$ for the isotope $^{45}$Sc. 
$Re\left[ \rho_{eg}\right] $ describes the refractive index from nuclei and results in the gravitational effects in our system as revealed by the last term of Eq.~\eqref{eq4}.
In cantrast, $Im\left[ \rho_{eg}\right]$ of Lorentzian line shape represents the nuclear absorption of x rays and leads to the Pound–Rebka experiment \cite{Pound1960}. 

In the following,  we use the SXWG   to simulate Rabi oscillations of x ray in a finite square well of width $L_x$.  
As illustrated in Fig.~\ref{fig1}(d),
we introduce a platinum cladding $n_e =1-\delta_\mathrm{Pt}+i\beta_\mathrm{Pt} $ for $\vert x\vert > L_x/2$  to constitute a finite square well potential, which provides with a transverse confinement to the x ray propagation direction \cite{Chen2022}.
The cladding material leads to the energy eigenfunctions of the $E$ field in Eq.~\eqref{eq4}  which read as (see supplemental information)
\begin{equation}
\psi_n\left( x\right) =\sqrt{\frac{2}{L_x}}\sin\left[ \frac{n\pi}{L_x}\left( x+\frac{L_x}{2}\right) \right], 
\end{equation}
with the eigen angular frequencies 
$
\omega_n = n^2\pi^2 c / \left( 2 k L_{x}^{2}\right) .
$
Inside the square well $\vert x\vert \leq L_x/2$,  we perturb the system by a periodic (gradient) particle distribution of isotope $X$ and carbon along the y direction (x direction), where $\xi \left( x,y\right) = \left( \varrho/2\right)  \left[ 1+\left( 2x/L_x\right)  \sin\left( k_d y\right) \right]$  and $n_e\left( x,y\right) =1-\delta_\mathrm{C}\left( x,y\right) +i\beta_\mathrm{C}\left( x,y\right)-\delta_\mathrm{X}\left( x,y\right)+i\beta_\mathrm{X}\left( x,y\right) $ (see supplemental information for the form of $n_e$).
In Fig.~\ref{fig1}(d)
subscripts Pt, C, and X represent platinum, carbon and resonant nucleus, respectively. 
We list other relevant material parameters in Table~\ref{table1}.
With the above density modulation, the last term in Eq.~\eqref{eq4} effectively becomes the electric dipole Hamiltonian in an oscillating field which perturbes the square well potential. This plays the key role to drive the x-ray Rabi oscillation with gravitational sensitivity.
When the resonant condition $c k_d = \omega_{n+1} - \omega_n$ is fulfilled,
the periodic structure of the refractive index  drives the dipole transition $\psi_n\rightarrow\psi_{n+1}$ with
the effective Rabi frequency (see supplemental information) 
\begin{equation}\label{Rabi}
\Omega_{n}=  \frac{16 \varrho n\left( n+1\right)  c}{\pi^2\left( 2n+1\right) ^2 L} \left[ 
\frac{k\left( \delta_\mathrm{C}-\delta_\mathrm{Sc}\right)}{2N\sigma_0 }+    \frac{2\Delta_t }{ 1+4\Delta_t^2}
\right].
\end{equation}
A propagating x ray experiences a constant SXWG-induced Rabi frequency, and the condition for having a  $m\pi$ pulse is  $\vert\Omega_{n}\vert L/c = m\pi$.
We use $m=2$ to demonstrate the Rabi oscillation between the x-ray ground state $\psi_1$ and the first excited state $\psi_2$ by numerically solving Eq.~\eqref{eq4}.

The solution of Eq.~\eqref{eq4} in Fig.~\ref{fig2} represents an x ray which propagates through an SXWG with
with $X=^{45}$Sc, $\varrho = 26.25$, natural scandium  particle density $N=3.99\times 10^{28}$m$^{-3}$, nuclear resonance absorption cross section $\sigma_0 = 12.6$(kbarn), $L_x=100$nm, $L=4$mm, $k_d=22.778\times 10^3$rad/m, $\Delta=4\Gamma$,  and $\Delta_G=0$.
The process is   visualized  in terms of the fidelity 
$F_{n}\left( y\right) =  \vert  \int_{-\infty}^\infty \psi_n^\ast\left( x\right)  E\left( x, y\right)  dx \vert^2 /  \int_{-\infty}^\infty \vert E\left( x, y\right) \vert^2 dx$ in Fig.~\ref{fig2}(a),   and  the normalized intra-waveguide x-ray intensity distribution $ \vert E\left( x, y\right) \vert^2  /  \int_{-\infty}^\infty \vert E\left( x, y\right) \vert^2 dx$ in Fig.~\ref{fig2}(b).
The alternation of $F_1$ and $F_2$ in Fig.~\ref{fig2}(a) clearly demonstrates that
the ground-state x ray  enters the SXWG  at $y=0$, and is then coherently promoted to the first excited state  when approaching $y=2$mm. 
After that, the x ray  returns to the ground state and finishes a full Rabi cycle at $y=4$mm. 
One can also observe the same phenomenon at the intra-waveguide x-ray intensity which evolves back and forth between states $\psi_1$ and $\psi_2$ in  Fig.\ref{fig2}(b).

%
%
The above x-ray Rabi oscillation suggests a systematic way to generate the $\psi_n$ mode with a sequence of SXWGs driving $\Delta n =1$  transitions, where one can raise the quantum number one by one.
Specifically,  one can connect two different SXWGs to accomplish  a $\psi_1 \rightarrow \psi_{3}$ transition, where the upstream SXWG drives a $\psi_1 \rightarrow \psi_{2}$ transition, and the downstream SXWG achieves a $\psi_2 \rightarrow \psi_{3}$ promotion.
Moreover, one can even design multiple SXWG modules for $\Delta n =2$ or any dipole forbidden transitions.
Thus, all combinations of SXWG modules open the capability to generate high order x-ray modes starting from the ground state $\psi_{1}$.
It is worth mentioning another possible application using  dual  SXWGs with a gap in between  as an x-ray  interferometer.
While the upstream SXWG causes the $\psi_1 \rightarrow \psi_2$ transition as a  beam splitter, the downstream SXWG leads to the return of $\psi_2 \rightarrow \psi_1$ as a beam combiner. 
Furthermore, in the gap between two SXWGs one can introduce a phase modulator to impose a phase shift at one branch of the split  state $\psi_2$, e.g.,   $ x>0$ at $y=2$mm in  Fig.~\ref{fig2}. A controllable interference  due to the  phase modulation is expected to happen at the end of the downstream SXWG.

We are ready to demonstrate the Earth's gravitational effect on our SXWG system. 
Given that the gravitational redshift $\Delta_G$ significantly changes the nuclear coherence $\rho_{eg}$ and Rabi frequency $\Omega_n$ in Eq.~\eqref{Rabi} in two millimeter  on Earth as demonstrated in Fig.~\ref{fig1}(c), 
this sensitivity  potentially allows for turning gravity into a practical use,  e.g.,  gravitationally sensitive  x-ray optics.  
For illustrating the effect, we numerically solve Eq.~\eqref{eq4} and use the isotope $^{45}$Sc  in an SXWG with parameters  $\varrho = 8.38$, $L_x=100$nm, $L=2$mm, $k_d=23.778\times 10^3$rad/m, and $\Delta=19.36$ to show the gravitational effect.
Fig.~\ref{fig3}(a) illustrates  three cases where the above discussed SXWG is located at $z=2.32$cm,  $z=2$cm,  and $z=1.72$cm from the top down. 
An incident x ray with the transverse mode $\psi_1$ is deflected upward and experiences a gravitational redshift (vertical upward arrow with color gradient). 
The $\Delta_G$ will change when the x ray illuminates the SXWG at different altitudes.   
The total detuning  for each case is specified at the level-scheme plot, namely, $\Delta+\Delta_G=-2.52$,  $\Delta+\Delta_G=0.5$,  and $\Delta+\Delta_G=3.14$ from the top down. 
We emphasize that the periodic particle density modulation effectively plays the role of a resonant field,  and it always resonantly drives a transition between the x-ray modes in a cladding waveguide for three cases. 
However, various $\Delta_G$ change the effective coupling strength $\Omega_n$ and result in different outputs.
The scattered/split x rays reflect the output mode and can be measured by a downstream position-sensitive detector. The $x$-dependent photon number counts show the output $\vert E\left( x, L\right) \vert^2$ and reveal the Earth's gravitational effect.
We depict the normalized  $\vert E\left( x,y\right) \vert^2$ for $z=2.32$cm,  $z=2$cm,  and $z=1.72$cm in Fig.~\ref{fig3}(c, e, and g), respectively.
The intra-waveguide intensity shows that the x ray significantly gets split in Fig.~\ref{fig3}(e) under a half Rabi cycle  $\psi_1 \rightarrow \psi_2$ in the SXWG at $z=2$cm. 
In contrast, Fig.~\ref{fig3}(c and g) depicts only a transverse broadening of the x ray  due to a small $\vert \Omega_1\vert$. 
Fig.~\ref{fig3}(b,d, and f) illustrate  $F_1\left( y\right)$ (red-dashed line) and $F_2\left( y\right)$ (green-solid line) for $z=2.32$cm,  $z=2$cm,  and $z=1.72$cm, respectively.
One can clearly see that the x ray experiences  Rabi flopping and becomes $\psi_2$ at the resonant altitude $z=2$cm as also pointed out by Fig.~\ref{fig3}(d and e).
Given that $\vert \Omega_1\vert$ decreases when the SXWG leaves the resonant altitude,   the x ray mostly remains in the initial mode $\psi_1$, namely, $F_1\left( y\right)>F_2\left( y\right)$ at $z=2.32$cm  and $z=1.72$cm.
We depict the output altitude-dependent $F_2$ at $y=2$mm in Fig.~\ref{fig4}(a).   
As a result,  different x-ray splitting is expected to occur when lifting an SXWG composed of $^{45}$Sc  at only a millimeter altitude change.

We quantify the gravitational sensitivity of the SXWG by the  full altitude width $\Delta Z_G$  at the half maximum $Re\left[ \rho_{eg}\right] $
\begin{equation}\label{zFWHM}
\Delta Z_G = \sqrt{3}\left( \frac{\hbar\Gamma}{E_t}\right)  \frac{ c^2 R_E^2}{ G M_E},
\end{equation}
as indicated by the black-horizontal double arrow in Fig.~\ref{fig1}(c).
The introduced $\Delta Z_G$ is a measure for  the sensitivity of the x-ray-nucleus coupling to the change of the SXWG vertical location.
With the definition of the quality factor of a nuclear resonance $Q=E_t/\left( \hbar\Gamma\right) $,  we can see that $\Delta Z_G$ is proportional to $1/Q$.
Fig.~\ref{fig4}(b) exemplifies the implication of Eq.~\eqref{zFWHM} for our system on  Earth in a double-logarithmic plot, where we mark the isotopes 
$^{45}$Sc,
$^{57}$Fe,  
$^{67}$Zn, 
$^{73}$Ge, 
$^{103}$Rh, 
$^{107}$Ag,
$^{109}$Ag,   
$^{181}$Ta, 
$^{182}$Ta, 
$^{229}$Th,
and
$^{249}$Bk, 
according their $Q$ factor. Some of the nuclear parameters are listed in Table~\ref{table1}. 
Remarkably,  the advantage of a very high $Q\sim 10^{19}$ of $^{45}$Sc or $^{182}$Ta nuclear resonance endues
an SXWG with a gravitational sensitivity to only millimeter altitude change. 
Notably, it is also possible to get 
sub-millimeter $\Delta Z_G$ using $^{229}$Th whose $Q\sim 10^{20}$ \cite{Seiferle2019}, and
micron  $\Delta Z_G$ with $^{107}$Ag and  $^{109}$Ag whose $Q\sim 10^{22}$. 
It deserves to mention that $^{103}$Rh  whose $Q > 10^{23}$ \cite{nndc} even results in a nanometer $\Delta Z_G$, and it may lead to gravitational application in mesoscopic scale.

In conclusion we have put forward a controllable  SXWG system that potentially turns gravity into an application of x-ray optics.
A periodic intra-waveguide structure, e.g., nuclear optical lattice $^{57}$Fe/$^{56}$Fe  bilayers in Ref.\cite{Haber2016}, can drive a transition between x-ray modes.
The x-ray transverse mode experiences Rabi oscillation when propagating in an SXWG.
Our scheme allows for applications like  a systematic production of structured x rays and an x-ray  interferometer without any beam splitter.
Remarkably, a significant change of a gravitionally induced splitting of x rays could happen by lifting our SXWG made of, e.g., $^{45}$Sc or $^{182}$Ta, at only a millimeter scale. 

S.-Y. L. and W.-T. L. are supported by the National Science and Technology Council of Taiwan (Grant No. 
110-2112-M-008-027-MY3, 
110-2639-M-007 -001-ASP,
111-2923-M-008-004-MY3 \&
111-2639-M-007-001-ASP).
S. A. is supported by National  Science Foundation of China (Grant No. 11975155).

\bibliography{20230309_NFSBS2A}

\end{document}